\documentclass[a4paper,11pt]{article}
\usepackage{graphicx}
\usepackage{amsmath}
\usepackage{amsbsy}
\begin{document}

\title{Gravitational radiospectrometer}

\author{G. S. Bisnovatyi-Kogan$^{1,2,3}$ and O. Yu. Tsupko$^{1,3}$
\\[5mm]
$^1$Space Research Institute of Russian Academy of Science,\\
Profsoyuznaya 84/32, Moscow 117997\\
$^2$Joint Institute for Nuclear Research, Dubna, Russia\\
$^3$Moscow Engineering Physics Institute, Moscow, Russia\\ \\
\it e-mail: gkogan@iki.rssi.ru, tsupko@iki.rssi.ru\\
}

\date{}
\maketitle \abstract{Gravitational lensing is predicted by general
relativity and is found in observations. When a gravitating body
is surrounded by a plasma, the lensing angle depends on a
frequency of the electromagnetic wave due to refraction
properties, and the dispersion properties of the light propagation
in plasma. The last effect leads to dependence, even in the
uniform plasma, of the lensing angle on the frequency, what
resembles the properties of the refractive prism spectrometer. The
strongest action of this spectrometer is for the frequencies
slightly exceeding the plasma frequency, what corresponds to very
long radiowaves. }

\section{Introduction}
An ordinary theory of the gravitational lensing is developed for
the light propagation in the vacuum. Gravi\-ta\-tio\-nal lensing
in the vacuum is achromatic because the deflection angle for the
photon does not depend on the frequency of the photon \cite{GL}.
In the limit of a weak lensing in vacuum by a body with a mass $M$
the deflection angle for the photon (Einstein angle) is
$\hat{\alpha} = 4GM/c^2 b = 2 r_g/b$, under condition $b \gg r_g$,
where $b$ is impact parameter, $r_g$ is the Schwarzschild radius
\cite{GL}.

Propagation of the light in the medium at presence of the gravity
field was considered by many authors \cite{Muhl1966},
\cite{Muhl1970}, \cite{Noonan}, \cite{B-Minakov} and references
there. If we consider inhomogeneous medium (without gravity) the
light rays move along the curved trajectories in this medium. In
the papers concerning gravitational deflection  the inhomogeneous
medium was considered. The deflection due to the gravitation, and
the deflection due to the inhomogenity of the medium had been
considered separately, without an account of the influence of the
dispersion in plasma on the light propagation in the gravitational
field. In this work we show that even in the homogeneous medium
the dispersion in the plasma leads to dependence of the light
deflection  angle on the wavelength, what is different from the
constant deflection angle in the vacuum.

It have been shown  \cite{GL}, \cite{B-Minakov}, \cite{Fok},
\cite{LL2}, \cite{Myoller}, that light propagation in the
gravitational field in the vacuum may be formulated as  its
propagation in a inhomogeneous medium with an effective refraction
index $n_g$, depending on metric. It have been shown also, that in
presence of the medium in the gravitation field, such analogy is
valid too. In this case we should use the refraction index, which is
a multiplication of the effective gravitational refraction index
$n_g$ and usual refraction index $n$, determined by the physical
properties of the medium: $n_{eff} = n_g n$ \cite{B-Minakov},
\cite{Noonan}. In the case when both $n_g$ and $n$ are close to
unity, the combined deviation of the refraction index from unity is
reduced to the sum of both separate effects \cite{Muhl1966},
\cite{Muhl1970}, \cite{B-Minakov}.

In this work we consider the gravitational lensing in a homogeneous
plasma. Plasma is a dispersive medium, where the refraction index
depends on the frequency of the photon. Therefore in the plasma the
photons with different frequencies move with different velocities,
namely the photons with smaller frequency (or bigger wavelength)
move with smaller group velocity of the light signal. We obtain
here, that in a homogeneous plasma, in the presence of gravity, the
deflection angle of the photon depends on the frequency of the
photon, and discuss observational effects of this phenomenon.

In the works, where the effect of the light dispersion in the plasma
was not taken into account, the dependence of the gravitational
deflection on frequency was connected only with the plasma
inhomogeneity, and disappeared in the uniform plasma.
 In the book of   Synge \cite{Synge}, the
geometrical optics in the medium with gravity was considered in
great details, and he had derived equations for the photon
propagation in an arbitrary medium with gravity. Here we
calculate, on the basis of equations of Synge \cite{Synge}, the
deflection angle for the photon wave packet, moving in the
gravitational field in the presence of a uniformly distributed
plasma. We obtain the dependence of the deflection angle on the
frequency for the case of a week field of the Schwarzschild black
hole metric. The deflection angle increases with decreasing of the
frequency, and when the frequency is approaching the electron
plasma frequency $\omega_e^2=\frac{4\pi e^2 n_e}{m_e}$, all
photons are falling into the black hole, if their impact parameter
is less than the critical one, depending on frequency, $b <
b_c(\omega)$. When $\omega$ approaching $\omega_e$, the critical
$b_c$ formally goes to $\infty$.

\section{Light propagation in an inhomogeneous plasma in a week gravitational field}

Let us consider a static space-time with the metric

\begin{equation}
ds^2 = g_{ik} \, dx^i dx^k = g_{\alpha \beta} \, dx^\alpha
dx^\beta + g_{00} \left(dx^0\right)^2 , \; \; i,k = 0,1,2,3, \; \;
\alpha, \beta = 1,2,3.
\end{equation}
Here $g_{ik}$  do not depend on the time. Let us assume that the
gravitational field is week, so that

\begin{equation}\label{metr}
g_{ik} = \eta_{ik} + h_{ik}, \; \; h_{ik} \ll 1, \; \; h_{ik}
\rightarrow 0 \; \; \mbox{under} \; \;  x^\alpha \rightarrow
\infty \, .
\end{equation}
Here $\eta_{ik}$ is the metric of a flat space  $(-1,1,1,1)$, and
$h_{ik}$ is a small perturbation. Note \cite{LL2}, that

\begin{equation}\label{metr1}
 g^{ik} =
\eta^{ik} - h^{ik}, \quad \eta^{ik}=\eta_{ik},\quad  h^{ik}=h_{ik}.
\end{equation}
Let us consider in this gravitational field a
static inhomogeneous plasma with the refraction index $n$, which depends
on the space location $x^\alpha$ and the frequency of the photon
$\omega(x^\alpha)$:
\begin{equation} \label{plasma-n}
n^2 = 1 - \frac{\omega_e^2}{[\omega(x^\alpha)]^2}, \quad \omega_e^2
= \frac{4 \pi e^2 N(x^\alpha)}{m}.
\end{equation}
Here $\omega(x^\alpha)$ is the frequency of the photon, which depends
on space coordinates $x^1$, $x^2$, $x^3$ due to presence of
gravitational field (gravitational red shift).  We denote
$\omega(\infty) = \omega$, and $e$ is the charge of the electron, $m$ is
the mass of the electron, $N(x^\alpha)$ is the electron
concentration in the inhomogeneous plasma, $\omega_e$ is the
electron plasma frequency in this plasma. To consider a general case
let us assume
\begin{equation} \label{plasmaN}
N(x^\alpha) = N_0 + N_1(x^\alpha),  \; \; N_0 = {\rm const}, \; \;
N_1(\infty) = 0.
\end{equation}
Here $N_1$ is not supposed to be small compared to  $N_0$, let us denote:

\begin{equation}
 \label{nonun}
\omega_e^2 = \omega_0^2 + \omega_1^2, \; \; \mbox{where} \; \;
\omega_0^2 = K_e N_0, \; \; \omega_1^2 = K_e N_1, \; \; K_e =
\frac{4 \pi e^2}{m}.
\end{equation}
The gravitational optic in a medium in a curved space-time, was
investigated in \cite{Synge}. Consider a set of three-dimensional
surfaces in a static four-D space-time, which are characterized by
monotonically increasing phase angle. These 3-D surfaces are
called 3-D waves, or phase waves. It was found in \cite{Synge},
the connection between the phase velocity $u$, and a 4-vector of
the photon momentum $p^i$, written with using the refraction index
of medium $n$, $n=c/u$, $c$ is the light velocity in a vacuum, as

\begin{equation} \label{eq-medium}
\frac{c^2}{u^2}=n^2 = 1 + \frac{p_i p^i}{\left(p^0
\sqrt{-g_{00}}\right)^2} \, .
\end{equation}
 The refraction index $n$, defined for plasma in (\ref{plasma-n}),
 is in a general case the function of $x^i$
and $\omega(x^\alpha)$, which are determined by the properties of the
medium, and the photon frequency. In the case of the vacuum ($n=1$) we
can obtain from (\ref{eq-medium}) the usual relation for the square
of the photon 4-vector: $p_i p^i = 0$. In the medium, the square of
the photon 4-vector is not equal to zero. For the medium in a flat
space-time we have,

$$g_{00}=-1, \,\,\, g_{\alpha\alpha}=1,\,\,\, p^0=-p_0,
\,\,\,p^\alpha=p_\alpha$$
\begin{equation}
n^2 = 1 + \frac{-(p^0)^2 + (p^\alpha)^2}{(p^0)^2},
\end{equation}
and obtain the usual relation between the space and time components
of the 4-vector of the photon  \cite{LL8}, \cite{Zhelezn},
\cite{Ginzb},

\begin{equation}
(p^\alpha)^2 = n^2 (p^0)^2.
\end{equation}
The time component of the photon 4-vector is its energy, therefore
$p^0$ is proportional to the frequency of the photon $\omega$ \cite{Myoller}.
We have in the flat space-time

\begin{equation}
p^0 = C \omega, \; \; C={\rm const}>0,
\end{equation}
and in a space-time with gravity we have
\cite{Myoller}, \cite{Synge},

\begin{equation} \label{intro-C}
p^0 \sqrt{-g_{00}} = C \, \omega(x^\alpha) \, ,
\end{equation}
what physically determines the gravitational red shift.
The coordinate $x^\alpha$ and the momentum $p^\alpha$
are connected by the relation
$d x^\alpha/d \lambda = p^\alpha$, where $\lambda$ is a parameter
changing along the photon trajectory (see below). Consider a
photon moving along $z$-axis, with the frequency at infinity equal to $\omega$.
Without the gravity and medium
inhomogeneity its unperturbed  trajectory is a straight line along $z$ axis.
The photon 4-vector for the unperturbed trajectory is
$p^i=(p^0,0,0,p^3)$,  and  the relation between the space and
time components, using the flat metric $\eta_{ik}$, is

\begin{equation}
(p^3)^2 = n_0^2 (p^0)^2 \, .
\end{equation}
Here we denote (\ref{nonun})

\begin{equation}
n_0 = n(\infty) = \sqrt{1 - \frac{\omega_0^2}{\omega^2}} .
\end{equation}
It is convenient to use the coordinate $z$ as the parameter
$\lambda$. Then we have the components of the photon 4-vector in
a simple form:

\begin{equation} \label{unpert-p}
 p^i = ( 1/n_0, 0, 0, 1 ) \, , \quad \, p_i = ( -1/n_0, 0, 0, 1 ) .
\end{equation}
Thus the photon momentum in plasma is a time-like 4-vector

\begin{equation}
p^i p_i =1 - \frac{1}{n_0^2}  =1 - \left( 1 -
\frac{\omega_0^2}{\omega^2} \right)^{-1} < 0 ,
\end{equation}
While the phase velocity $u=\frac{c}{n_0}$ in plasma is larger than the light velocity in
vacuum $c$,
the group velocity $v_{gr}$ is less than $c$,
so that the larger frequency corresponds to the larger group velocity, tending to
$c$ in the limit. For the group
velocity we have the relation \cite{LL8}, \cite{Synge}

\begin{equation}
\frac{c}{v_{gr}} = \left| \frac{\partial}{\partial \omega} (n_0
\omega) \right| = \frac{1}{\sqrt{1 - \omega_0^2 / \omega^2}} \, ,
\quad v_{gr} = c \sqrt{1 - \frac{\omega_0^2}{\omega^2}}=c\,n_0 < c \, ,
\end{equation}
so that $ v_{gr} u=c^2$. To find a constant $C$ in (\ref{intro-C}), we consider
this relation at infinity, where $\omega(x^\alpha) =\omega$, and $p^0$ is defined by
 (\ref{unpert-p}). We have then

\begin{equation} \label{const}
p^0 = C \omega, \qquad C = \frac{1}{n_0 \omega}.
\end{equation}
The trajectories of the photon in presence of the gravitational field may
be found from the variational principle \cite{Synge}

\begin{equation} \label{var-princ}
\delta \left(\int p_i \, dx^i\right) = 0 ,
\end{equation}
with the restriction  (\ref{eq-medium}), which may be written in
 the form

\begin{equation} \label{add-cond}
W(x^i,p_i) = \frac{1}{2} \left[ g^{ij} p_i p_j - (n^2-1) \left(p_0
\sqrt{-g^{00}}\right)^2 \right] = 0.
\end{equation}
Here we define the scalar function $W(x^i,p_i)$ of
$x^i$ and  $p_i$.
The variational principle (\ref{var-princ}), with the restriction
condition $W(x^i,p_i)=0$, leads to the system of differential
equations \cite{Synge}:

\begin{equation}
\label{D-Eq} \frac{dx^i}{d \lambda} = \frac{\partial W}{\partial
p_i} \, , \; \; \frac{dp_i}{d \lambda} = - \frac{\partial
W}{\partial x^i} \, ,
\end{equation}
with the parameter $\lambda$  changing along the light trajectory.
Let us introduce a variable

\begin{equation} \label{chi}
 \chi = p_0 \sqrt{-g^{00}} =-p^0 \sqrt{-g_{00}} = - C \,
\omega(x^\alpha) = - \frac{1}{n_0 \omega} \, \omega(x^\alpha),
\end{equation}
by using of which we can transform $W(x^i,p_i)$ to a  simpler
form

\begin{equation} \label{W-alpha}
W(x^i,p_i) =  \frac{1}{2} \left[ g^{ij} p_i p_j - (n^2-1)
\left(p_0 \sqrt{-g^{00}}\right)^2 \right] =
\end{equation}
$$
= \frac{1}{2} \left[ g^{00} p_0 p_0 + g^{\alpha \beta} p_\alpha
p_\beta - (n^2-1) p_0^2 (-g^{00}) \right] = \frac{1}{2} \left[
g^{\alpha \beta} p_\alpha p_\beta - n^2 \chi^2 \right] .
$$
From (\ref{D-Eq}) we obtain the system of equations for the space
components $p_\alpha$:

\begin{equation} \label{syst-gen}
\frac{dx^\alpha}{d \lambda} = g^{\alpha \beta} p_\beta, \quad
\frac{dp_\alpha}{d \lambda} = - \frac{1}{2} \, g^{\beta \gamma}_{,
\alpha} p_\beta p_\gamma + \frac{1}{2} \left(n^2 \chi^2\right)_{,
\alpha} .
\end{equation}
For the inhomogeneous plasma $n=n(\chi, x^\alpha)$, and we have, using
(\ref{plasma-n}),(\ref{plasmaN}), (\ref{chi}) the relation
\begin{equation}
\frac{1}{2} \left(n^2 \chi^2\right)_{, \alpha} = \frac{1}{2}
\left[ \left( 1 - \frac{K_e N_0}{[\omega(x^\alpha)]^2} - \frac{K_e
N_1(x^\alpha)}{[\omega(x^\alpha)]^2}   \right) \chi^2 \right]_{,
\alpha} =
\end{equation}
$$
= \frac{1}{2} \left[ \chi^2 - \frac{1}{n_0^2 \omega^2} K_e N_0 -
\frac{1}{n_0^2 \omega^2} K_e N_1(x^\alpha) \right]_{, \alpha} =
\chi\frac{\partial \chi}{\partial x^\alpha} - \frac{1}{2} \,
\frac{K_e}{n_0^2 \omega^2}\frac{\partial
N_1(x^\alpha)}{\partial x^\alpha} .
$$
As follows from (\ref{D-Eq}), the variable $p_0$ is constant along
the trajectory, so from (\ref{chi}) we have

\begin{equation}
\frac{\partial \chi}{\partial x^\alpha} = p_0 \left(
\sqrt{-g^{00}} \right)_{, \alpha} ,
\end{equation}
and obtain finally
\begin{equation}\label{chider}
\frac{1}{2} \left(n^2 \chi^2\right)_{, \alpha} = \chi \, p_0
\left( \sqrt{-g^{00}} \right)_{, \alpha} - \frac{1}{2} \,
\frac{1}{n_0^2 \omega^2} K_e \frac{\partial
N_1(x^\alpha)}{\partial x^\alpha} = \frac{1}{2} \, p_0^2 \left(
-g^{00} \right)_{, \alpha} - \frac{1}{2} \,
\frac{1}{\omega^2-\omega_0^2} \, K_e \frac{\partial
N_1(x^\alpha)}{\partial x^\alpha} .
\end{equation}
Using (\ref{chider}), we reduce equations (\ref{syst-gen}) to the form

\begin{equation} \label{Syst}
\frac{dx^\alpha}{d \lambda} = g^{\alpha \beta} p_\beta, \quad
\frac{dp_\alpha}{d \lambda} = - \frac{1}{2} \, g^{\beta \gamma}_{,
\alpha} p_\beta p_\gamma + \frac{1}{2} \, p_0^2 \left( -g^{00}
\right)_{, \alpha} - \frac{1}{2} \, \frac{1}{\omega^2-\omega_0^2}
\, K_e \frac{\partial N_1(x^\alpha)}{\partial x^\alpha} \, .
\end{equation}
The system (\ref{Syst}) describes the light propagation in
an inhomogeneous plasma, with account the dispersion,
in the presence of a gravitational field.
Let us consider the photon moving along $z$-axis in a flat space with a
homogeneous plasma, and use the
coordinate $z$ as the parameter $\lambda$. In approximation of
small perturbation $h_{ik}$ and week inhomogeneity,
$N_1\ll N_0$, one can integrate equations ,
calculating right-hand side of equations (\ref{Syst}) by using
the unperturbed
trajectory of the photon, with $p^i$ from (\ref{unpert-p}), in the right hand side.
At this simplification, the first and second terms in the second equation
 (\ref{Syst}) can be written as

\begin{equation}
 - \frac{1}{2} \,
g^{\beta \gamma}_{, \alpha} p_\beta p_\gamma + \frac{1}{2} \,
p_0^2 \left( -g^{00} \right)_{, \alpha} = - \frac{1}{2} \,
 g^{33}_{, \alpha} p_3^2 + \frac{1}{2} \, p_0^2 \left( -g^{00}
\right)_{, \alpha} \, .
\end{equation}
In the week gravitational field we can write, using
(\ref{metr}),(\ref{metr1}) and (\ref{unpert-p})

\begin{equation}
- \frac{1}{2} \, g^{33}_{, \alpha} p_3^2 + \frac{1}{2} \, p_0^2
\left( -g^{00} \right)_{, \alpha} =  \frac{1}{2} \, (h_{33})_{,
\alpha} + \frac{1}{2} \, \frac{1}{n_0^2} \left( h_{00} \right)_{,
\alpha}  = \frac{1}{2} \left( h_{33} + \frac{\omega^2\, h_{00}}
{\omega^2 - \omega_0^2} \right)_{, \alpha} \, .
\end{equation}
Finally we obtain the equation describing the light propagation
in a weekly inhomogeneous plasma, in the presence of a small
gravitational field, as

\begin{equation}\label{eq}
\frac{dp_\alpha}{d z} = \frac{1}{2} \left( h_{33} +
\frac{\omega^2\, h_{00}}
{\omega^2 - \omega_0^2}
 \right)_{, \alpha}  - \frac{1}{2} \,
\frac{K_e}{\omega^2 - \omega_0^2} \, \frac{\partial
N_1(x^\alpha)}{\partial x^\alpha}  \, .
\end{equation}
The deflection angle in the plane, perpendicular to the unperturbed
light trajectory, is determined as

\begin{equation}\label{angl}
\hat{\alpha}_\beta = [p_\beta(+\infty) -
p_\beta(-\infty)]/p, \quad p=\sqrt{p_1^2+p_2^2+p_3^2}=|p_3|=1,\quad
\beta=1,2.
\end{equation}
Here $p$ is defined by the unperturbed trajectory, and $\beta=1,2$ are related to
$x,y$ axes.
After integration, we have the following expression for the deflection angle of
the photon with the unperturbed trajectory
along  $z$ axis

\begin{equation}
\hat{\alpha}_\beta = \frac{1}{2} \int \limits_{-\infty}^{\infty}
\frac{\partial}{\partial x^\beta} \left( h_{33}
+\frac{h_{00}\omega^2}{\omega^2 -
\omega_0^2}
-\frac{K_e N_1(x^\alpha)}{\omega^2 - \omega_0^2} \,
\right) dz .
\end{equation}
 If the problem is axially symmetric, it is convenient to introduce the
impact parameter $b$.
Let us consider the photon moving along the axis $z$ in the plasma with
the impact parameter $b$ relative to the point mass, and the plasma has
a spherically-symmetric distribution of a concentration around this
point mass, $N = N(r) = N_0 + N_1(r)$. The situation is axially
symmetric, so the position of the photon is given by $b$ and $z$,
and the absolute value of the 3-radius-vector is $r = \sqrt{b^2+z^2}$,
instead of $r = \sqrt{x_1^2+x_2^2+z^2}$. We have the following
expression for the deflection angle in the plane perpendicular to
direction of the unperturbed photon:

\begin{equation}
\label{angle}
\hat{\alpha}_b = \frac{1}{2} \int \limits_{-\infty}^{\infty}
\left( \frac{\partial h_{33}}{\partial b}  +  \frac{\omega^2}{\omega^2 -
\omega_0^2} \frac{\partial h_{00}}{\partial b}  -
\frac{K_e}{\omega^2 - \omega_0^2}  \frac{\partial
N_1(r)}{\partial b}   \right) dz .
\end{equation}
Note that $\hat{\alpha}_b < 0$ corresponds to bending of the light
trajectory to the direction of the gravitation center, and
$\hat{\alpha}_b > 0$ corresponds to the opposite deflection.

\section{Particular cases}

\subsection{Vacuum, and a homogeneous medium without dispersion}

Let us consider a homogeneous medium without dispersion, with
the refraction index $n$=const $\ge 1$ not depending on the frequency $\omega$.
The equations describing the light propagation follow from (\ref{syst-gen}):

\begin{equation}
\frac{dx^\alpha}{d \lambda} = g^{\alpha \beta} p_\beta, \quad
\frac{dp_\alpha}{d \lambda} = - \frac{1}{2} \, g^{\beta \gamma}_{,
\alpha} p_\beta p_\gamma + \frac{n^2}{2} \left(\chi^2\right)_{,
\alpha}
\end{equation}
The right-hand side of equation (\ref{syst-gen}) for
the $p_\alpha$ is transformed, using (\ref{chi}),(\ref{unpert-p}), and
for determination of the deflection angle $\hat{\alpha}_b$ of the photon,
moving with the impact parameter $b$ in the spherically symmetric
gravitational field, we have

\begin{equation}
- \frac{1}{2} \, g^{\beta \gamma}_{, \alpha} p_\beta p_\gamma +
\frac{n^2}{2} \left(\chi^2\right)_{, \alpha} = - \frac{1}{2} \,
g^{\beta \gamma}_{, \alpha} p_\beta p_\gamma + \frac{n^2}{2}
\, p_0^2 \left( -g^{00} \right)_{, \alpha} = - \frac{1}{2} \,
 g^{33}_{, \alpha} p_3^2 + \frac{n^2}{2} \, p_0^2 \left( -g^{00}
\right)_{, \alpha} =
\end{equation}
$$
= \frac{1}{2} \, (h_{33})_{, \alpha} \, p_3^2 + \frac{n^2}{2}
\, p_0^2 \left( h_{00} \right)_{, \alpha} = \frac{1}{2} \,
(h_{33})_{, \alpha} + \frac{n^2}{2} \, \frac{1}{n^2} \left(
h_{00} \right)_{, \alpha} = \frac{1}{2} \,
\frac{\partial}{\partial x^\alpha} \left( h_{00} + h_{33} \right)
$$
After integration we obtain the expression for the deflection
angle of the photon which coincides with the well known formula
\cite{BK&Ts}, \cite{DamEsp}, \cite{Far1992} for the vacuum case,
derived from the geodesic equation,

\begin{equation} \label{GravCosm}
\hat{\alpha}_\beta = \frac{1}{2} \int \limits_{-\infty}^{\infty}
\frac{\partial}{\partial x^\beta} \left( h_{33} + h_{00} \right)
dz .
\end{equation}
We see here, that in case of the non-dispersive medium, the constant index of
the refraction is canceled, and the photon trajectory  is
the same as in the vacuum, in presence of the gravitational field, in spite of
lower velocity of the light propagation in the medium. Note that
the motion of photons in 4-space, in the medium, is not described by the geodesic equation
(neither massive not zero), because the light propagation in the medium
with the refraction is determined not only by the gravitational field,
but also by the medium.

\subsection{Homogeneous plasma in a week gravitational field}

Here we represent the main new result of this
work:  the dependence of the deflection angle on the photon frequency in the
homogeneous plasma, due to account of the dispersion.
In the homogeneous plasma from (\ref{angle}), we have expression for the deflection
angle as
\begin{equation}\label{inthom}
\hat{\alpha}_b = \int \limits_{0}^{\infty}
\frac{\partial}{\partial b} \left( h_{33} + \frac{1}{1 -
\omega_0^2 / \omega^2} h_{00} \right) dz .
\end{equation}
Let us calculate the deflection angle for the photon moving in the
homogeneous plasma in the Schwarzschild metric \cite{LL2} of the
point mass $M$

\begin{equation}
ds^2 = - (1-r_g/r) \, dt^2 + \frac{dr^2}{1-r_g/r} + r^2(d \theta^2
+ \sin^2 \theta \, d\varphi^2),
\end{equation}
where $r_g=\frac{2GM}{c^2}$ is the Schwarzschild gravitational radius.
In the week field approximation this metric is written as \cite{LL2}

\begin{equation}
ds^2 = ds_0^2 + \frac{r_g}{r} (c^2 dt^2 + dr^2),
\end{equation}
where $ds_0^2$ is the flat part of metric ($ds_0^2 = \eta_{ik}
dx^i dx^k$). The components  $h_{ik}$ are written in the Cartesian
frame as \cite{LL2}

\begin{equation}
h_{00} = \frac{r_g}{r}, \quad h_{\alpha \beta} = \frac{r_g}{r}
n_\alpha n_\beta, \quad h_{33} = \frac{r_g}{r} \cos^2 \theta .
\end{equation}
Here $n_\alpha$ is the unit vector of 3-radius-vector $r_\alpha =
(x_1,x_2,x_3)$, the angle $\theta$ is the polar angle
between 3-vector $r^\alpha$ and $z$-axis, and $\cos \theta = z/r =
z/\sqrt{b^2+z^2}$. The integrals in (\ref{inthom}) are taken analytically:

\begin{equation}
\int \frac{\partial}{\partial b} h_{00} \, dz = \int
\frac{\partial}{\partial b} \frac{r_g}{\sqrt{b^2+z^2}} \, dz = -
r_g \int \frac{b}{(b^2+z^2)^{3/2}} \, dz = - r_g \, \frac{z}{b \,
\sqrt{b^2+z^2}} + {\rm const},
\end{equation}
\begin{equation}
\int \frac{\partial}{\partial b} h_{33} \, dz = \int
\frac{\partial}{\partial b} \frac{r_g}{\sqrt{b^2+z^2}}
\frac{z^2}{b^2+z^2} \, dz = - r_g \int \frac{3 z^2
b}{(b^2+z^2)^{5/2}} \, dz = - r_g \, \frac{z^3}{b \,
(b^2+z^2)^{3/2}} + {\rm const}.
\end{equation}
Substituting the limits of integration, we obtain the following
expression for the deflection angle in the homogeneous plasma

\begin{equation}\label{defl}
\hat{\alpha}_b = \int \limits_{0}^{\infty}
\frac{\partial}{\partial b} \left( h_{33} + \frac{1}{1 -
\omega_0^2 / \omega^2} h_{00} \right) dz = - \frac{r_g}{b} \left(
1 + \frac{1}{1 - \omega_0^2 / \omega^2} \right) .
\end{equation}
Here $\hat{\alpha}_b < 0$ for $\omega> \omega_0$, what means that the light ray is
bent to the direction of gravitation center, as it occurs in the
vacuum. The formula  (\ref{defl}) is valid only for $\omega> \omega_0$, because
the waves with $\omega< \omega_0$ do not propagate in the plasma \cite{Ginzb}.
In the theory of the gravitational lensing the deflection angle is usually
defined as the difference between the initial and the final ray directions
$\boldsymbol{\hat{\alpha}} = \textbf{e}_{in} - \textbf{e}_{out}$,
where $\textbf{e}$ is the unit tangent vector of a ray \cite{GL}.
Therefore, if we use this definition, we will have the expression with
the opposite sign:
\begin{equation} \label{main-res}
\hat{\alpha} = \frac{r_g}{b} \left( 1 + \frac{1}{1 - \omega_0^2 /
\omega^2} \right),
\end{equation}
which turns into the deflection angle \cite{GL} for vacuum
$2r_g/b$, when $\omega \rightarrow \infty$. For lower frequencies
the deflection angle may be much larger than in the vacuum, and
the image of the point source will be represented by the line, or
the ring (see Fig.1,2), on which the frequency is decreasing with
increasing of the distance from the source on the plane of the
view. Such effect may happen only for photons with radio
frequencies, because optical frequencies are much higher than the
plasma frequency $\omega_0$, so the effect should be negligible.
We see therefore, that the gravitational lens in plasma is acting
as {\bf the gravitational radiospectrometer}, see Fig.1,2.

\subsection{Inhomogeneous plasma with gravity}

Let us consider inhomogeneous plasma in presence of a week gravitational
field. Let us consider a case when
plasma concentration $N(\infty) =0$. In our notations we have $N_0
= 0$, $\omega_0=0$, $N = N_1$. In this case we have from (\ref{angle})
the following expression for the
deflection angle, valid when $|n-1|\ll 1$,
\begin{equation} \label{angle1-BM}
\hat{\alpha}_b = \frac{1}{2} \int \limits_{-\infty}^{\infty}
\left( \frac{\partial h_{33}}{\partial b}  +  \frac{\partial
h_{00}}{\partial b}  - \frac{1}{\omega^2} \, K_e \frac{\partial
N_1(r)}{\partial b} \right) dz .
\end{equation}
The case of the inhomogeneous plasma with gravity was considered in
\cite{Muhl1966}, \cite{Muhl1970}, \cite{B-Minakov}. Let us
calculate the deflection angle for the photon, moving in the
Schwarzschild metric, with the concentration of plasma in the
form \cite{B-Minakov}

\begin{equation} \label{N-BM}
N(r) = N_m \left( \frac{R}{r} \right)^h, \; \; N_m = {\rm const}, \; \;
R = {\rm const}, \; \; h>0 .
\end{equation}
It is evident from (\ref{angle1-BM}) that in the week
gravitational field, in presence of plasma which implies a small
perturbation in the photon propagation $(\omega^2 \gg
\omega_e^2)$, both effects: the deflection due to gravity and the
deflection due to inhomogeneity of medium (non-relativistic
effect) can be considered separately. The first and the second
terms in (\ref{angle1-BM}) lead to usual Einstein angle of
deflection, following from (\ref{main-res}) at $\omega_0^2=0$. Let
us calculate deflection due to the inhomogeneity of plasma, in the
third term in (\ref{angle1-BM}) for the density distribution
(\ref{N-BM}):

\begin{equation}
\frac{1}{2} \int \limits_{-\infty}^{\infty} \left( -
\frac{1}{\omega^2} \, K_e \frac{\partial N_1(r)}{\partial b}
\right) dz  =  - \frac{1}{\omega^2} \, K_e N_m \, R^h \int
\limits_{0}^{\infty} \frac{\partial}{\partial b} \left(
\frac{1}{r^h} \right) dz \, .
\end{equation}
After differentiation

\begin{equation}
\frac{\partial}{\partial b} \left( \frac{1}{r^h} \right) =
\frac{\partial}{\partial b} \frac{1}{(z^2+b^2)^{h/2}}  = - \frac{h
b}{(z^2+b^2)^{h/2+1}}
\end{equation}
we perform the integration using \cite{GR}, and properties of the
$\Gamma$-function:

\begin{equation}
\int \limits_{0}^{\infty} \frac{dz}{(z^2+b^2)^{h/2+1}} =
\frac{1}{h b^{h+1}} \frac{\sqrt{\pi} \, \Gamma\left(\frac{h}{2} +
\frac{1}{2}\right)}{\Gamma\left(\frac{h}{2}\right)} , \; \;
 \Gamma(x) = \int \limits_0^\infty t^{x-1} e^{-t} dt
.
\end{equation}
So we obtain

\begin{equation}
\int \limits_{0}^{\infty} \frac{\partial}{\partial b}
\frac{1}{(z^2+b^2)^{h/2}} dz = - \frac{1}{b^h} \frac{\sqrt{\pi} \,
\Gamma\left(\frac{h}{2} +
\frac{1}{2}\right)}{\Gamma\left(\frac{h}{2}\right)} \, ,
\end{equation}
and for the third term in (\ref{angle1-BM}) we have

\begin{equation}
\frac{1}{2} \int \limits_{-\infty}^{\infty} \left( -
\frac{1}{\omega^2} \, K_e \frac{\partial N_1(r)}{\partial b}
\right) dz  =  \frac{1}{\omega^2} \frac{4 \pi e^2}{m} N_m
\left(\frac{R}{b}\right)^h \frac{\sqrt{\pi} \,
\Gamma\left(\frac{h}{2} +
\frac{1}{2}\right)}{\Gamma\left(\frac{h}{2}\right)} \, .
\end{equation}
Finally, for the deflection angle near the Schwarzschild metric
with plasma concentration (\ref{N-BM}) we obtain the expression

\begin{equation} \label{angle-fin-BM}
\hat{\alpha}_b = -\frac{r_g}{b} + \frac{1}{\omega^2} \frac{4 \pi
e^2}{m} N_m \left(\frac{R}{b}\right)^h \frac{\sqrt{\pi} \,
\Gamma\left(\frac{h}{2} +
\frac{1}{2}\right)}{\Gamma\left(\frac{h}{2}\right)} \, .
\end{equation}
which is the same, as in \cite{B-Minakov}, where this
formula is written in terms of the wave length
instead of the frequency.

\section{Discussion}

Let us consider the case of a weekly inhomogeneous medium without
gravity, with the refraction index $n = n_0 + n_1$, $n_0 =$const,
$n_1 \ll n_0$. The system of equation for the trajectory of the
photon in this case follows from equations (\ref{syst-gen}) with
the flat metric $g_{ik}=\eta_{ik}$, see \cite{LL8},

\begin{equation} \label{inhom-n}
\frac{d x^\alpha}{dz} = p^\alpha, \; \; \frac{d p^\alpha}{dz} = -
\frac{1}{2} \, \eta^{\beta \gamma}_{, \alpha} p_\beta p_\gamma +
\frac{1}{2} \left(n^2 \chi^2\right)_{, \alpha} = \frac{1}{2 n_0^2}
\frac{\partial n^2}{\partial x^{\alpha}} \simeq \frac{1}{n_0}
\frac{\partial n}{\partial x^{\alpha}} \, .
\end{equation}
The light propagation in a week gravitational field may be
considered, as a propagation in a flat space with the
"gravitational" refraction index, which for the Schwarzschild
metric is written as \cite{GL, Muhl1966, Muhl1970, Noonan,
B-Minakov, Fok}

\begin{equation} \label{ng}
n_g=1+\frac{r_g}{r}, \quad \frac{r_g}{r} \ll 1.
\end{equation}
The total effective refraction index $n_{eff}$,
in presence of plasma with the proper refraction
index $n$, is determined
\cite{Noonan},\cite{B-Minakov} as $n_{eff}=nn_g$.
When plasma in a week gravitational field has a refraction index close to unity,
it follows from the definition of $n_{eff}$, that
in the linear approximation the effective
refraction index is obtained as a sum of two different additions to the unity
\cite{Muhl1966}, \cite{Muhl1970},\cite{B-Minakov}:

\begin{equation}\label{ntot}
n_{eff} = 1 + \frac{r_g}{r} - \frac{\omega_e^2(r)}{2\omega^2}, \;
\; \frac{r_g}{r} \ll 1, \; \; \frac{\omega_e^2(r)}{\omega^2} \ll
1.
\end{equation}
From (\ref{inhom-n}) we obtain
the deflection angle (\ref{angle-fin-BM}).
Note that when $|n^2-1|$ is not small, $n_{eff}$ cannot be represented in
the form (\ref{ntot}), and such case was considered in the subsection 3.2.

The main new result of this work is obtaining of the dependence,
of the lensing angle on the frequency in a
homogeneous plasma in the gravitational field. This effect has a relativistic
nature, and is connected with the dispersive properties of plasma.
It is interesting, that, as shown in the subsection 3.1,
in the medium without dispersion, the trajectories of photons
with different frequencies (energies) are exactly the same as in the vacuum, while
their velocities are less that the vacuum light velocity $c$.

Observational effect of such frequency dependence is easy to
explain on the example of lensing by the Schwarzschild point-mass
lens. This lens gives two images of source, on the opposite sides
from lens. Angular positions of images depend on the Schwarzschild
radius of lens and positions of source, lens and observer. The
dependence of the deflection angle on the frequency in plasma lead
to the phenomenon, that instead of two concentrated images with
complicated spectra, we will have two line images, formed by the
photons with different frequencies, which are deflected by
different angles (Fig.1,2).

The description of the mass distribution as a point mass
(Schwarzschild lens) is rarely sufficient for gravitational
lensing considerations \cite{GL}. In reality the
gravitational lenses have more a complicated structure, and
position of images different from that of the point-mass lens.
But all standard
models of gravitational lenses are based on the same Einstein deflection angle
$2r_g/b$, which should be modified for sufficiently long waves, according
to our formula (\ref{main-res}), in the presence of plasma. Note also, that taking into
account of
plasma effects on the gravitational lensing may influence the spectrum of the
microwave background radiation, leading to the dependence of power spectrum of the
fluctuations on the photon wavelength.

The light signals are propagated with the group velocity. It
follows from (\ref{main-res}), that the smaller group velocity
(smaller frequency and bigger wavelength) corresponds to a larger
deflection angle. Hence the effect of difference in the
gravitational deflection angles is significant for larger
wavelengths, when $\omega$ is approaching $\omega_e$, what is
possible only for the radio waves. Therefore, the gravitating
center in plasma is acting as a radiospectrometer. The longest
radiowaves are registered \cite{Braude} in the band $\sim 3 \cdot
10^3$ cm, corresponding to $\nu \simeq 10^7$ Hz $= 10$ MHz and
$\omega = 2 \pi \nu \simeq 6 \cdot 10^7$ sec$^{-1}$. The
spectroscopic effects of lensing will be important when $N_e \geq
3 \cdot 10^5$ cm$^{-3}$, corresponding to 10 $\%$ difference in
the lensing angles. Such electron densities may be expected around
the supermassive black holes, or during lensing at earlier stages
of the universe expansion at $z \geq 10^3$.

\begin{figure}
\centerline{\hbox{\includegraphics[width=0.6\textwidth]{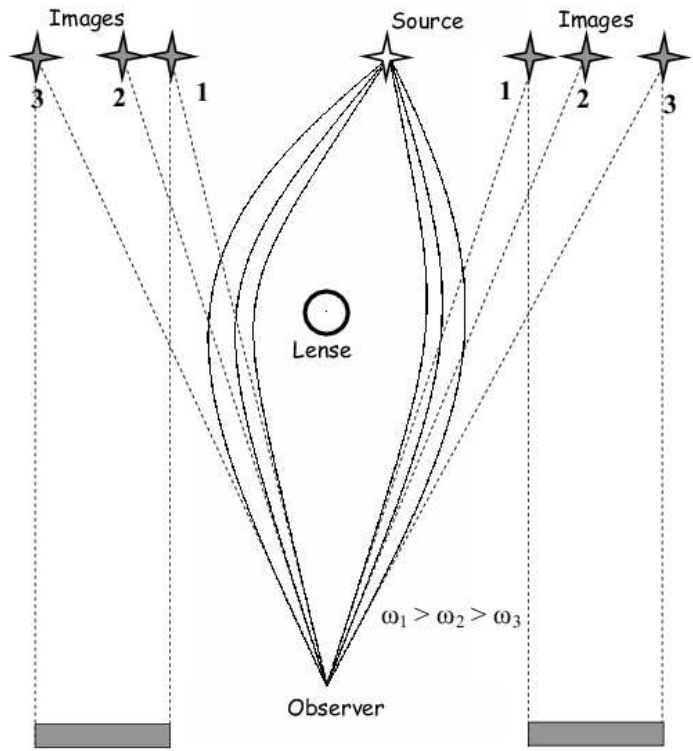}}}
\caption{Lensing of the point source by the Schwarzschild
point-mass lens. Instead of two point images due to lensing in the
vacuum we have two line images. The pairs of images, corresponding
to the same photon frequency, are indicated by the same numbers.
Two images with number 1 correspond to the vacuum lensing. }
\end{figure}

\begin{figure}
\centerline{\hbox{\includegraphics[width=0.6\textwidth]{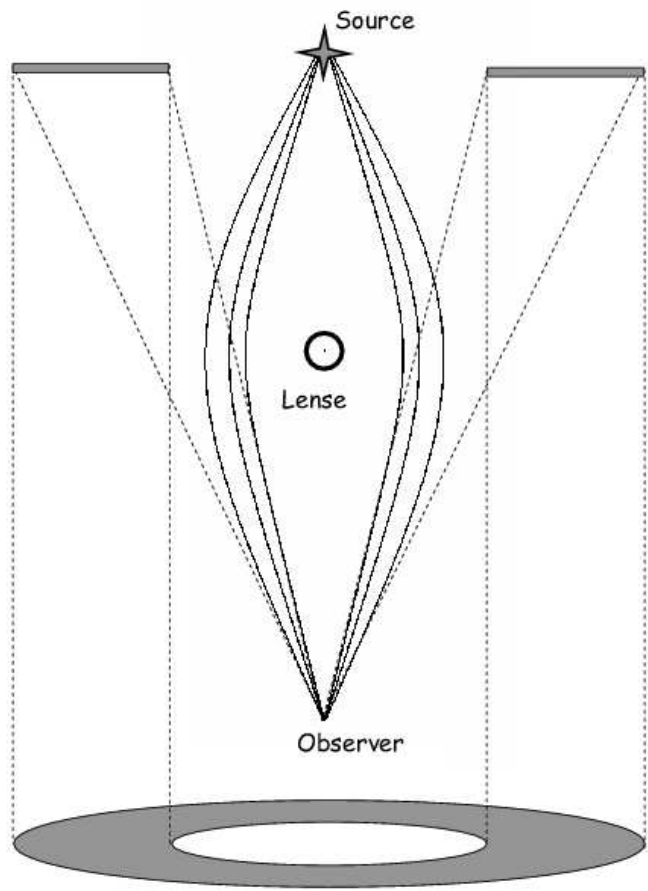}}}
\caption{Axis on lensing by the Schwarzschild point-mass lens. The
case of the Einstein ring. Instead of a thin ring corresponding to
the vacuum lensing (the inner circle of the ring) we have a thick
ring, formed by the photons of different frequencies.}
\end{figure}


\begin{thebibliography}{99}


\bibitem{GL}
Schneider, P.; Ehlers, J.; Falco, E.; Gravitational lensing,
Springer-Verlag, Berlin, 1992.

\bibitem{Muhl1966}
Muhleman, D.O.; Johnston, I.D.; Radio propagation in the solar
gravitational field, Phys. Rev. Lett. \textbf{17}, 8, 455 (1966).

\bibitem{Muhl1970}
Muhleman, D.O.; Ekers, R.D.; Fomalont, E.B.; Radio interferometric
test of the general relativistic light bending near the Sun; Phys.
Rev. Lett. \textbf{24}, 24, 1377 (1970).

\bibitem{Noonan}
Noonan, T.W.; Light rays in gravitating, refractive media; ApJ,
\textbf{262}, 344, 1982.

\bibitem{B-Minakov}
Bliokh, P.V.; Minakov, A.A.; Gravitational Lenses, Kiev, Naukova
Dumka, 1989 (in Russian).


\bibitem{Fok}
Fok, V.A.; Theory of Space, Time, and Gravitation, Moscow, 1955
(in Russian).


\bibitem{LL2}
Landau, L.D.; Lifshitz, E.M.; The Classical Theory of Fields,
Pergamon, Oxford, 1993.


\bibitem{Myoller}
M\o ller, C.; The Theory of Relativity, Clarendon Press, Oxford,
1972.


\bibitem{Synge}
J.L. Synge, Relativity: the General Theory, North-Holland
Publishing Company, Amsterdam, 1960.


\bibitem{LL8}
Landau, L.D.; Lifshitz, E.M.; Electrodynamics of Continuous Media,
Course of theoretical physics, Oxford, Pergamon Press, 1960.

\bibitem{Zhelezn}
V.V. Zhelezniakov, Electromagnetic Waves in Space Plasma:
Generation and Propagation, Moscow, Nauka, 1977(in Russian).

\bibitem{Ginzb}
Ginzburg, V.L.; The Propagation of Electromagnetic Waves in
Plasmas, International Series of Monographs in Electromagnetic
Waves, Oxford, Pergamon, 1970.

\bibitem{BK&Ts}
Bisnovatyi-Kogan, G.S.; Tsupko, O.Yu.; Gravitational lensing by
gravitational waves, Gravitation and Cosmology, \textbf{14}, 226,
2008.


\bibitem{DamEsp}
Damour, T.; Esposito-Far\`{e}se, G.; Light deflection by
gravitational waves from localized sources; Phys. Rev. \textbf{D}
\textbf{58}, 042001 (1998).

\bibitem{Far1992}
Faraoni, V.; Nonstationary gravitational lenses and the Fermat
principle; Astrophys. J. \textbf{398}, 425 (1992).


\bibitem{GR}
Gradshtein, I.S.; Ryzhik, I.M.; Tables of Integrals, Sums, and
Products, Nauka, Moscow, 1971 (in Russian).


\bibitem{Braude}
Braude, S.Ya.; Megn, A.V.; Rashokovski, S.L.; Ryabov, B.P.;
Sharykin, N.K.; Sokolov, K.P.; Tkatchenko, A.P.; Zhouk, I.N.; The
UTR-2 Very Low-Frequency Sky Survey Data (Braude+ 1978-2002),
VizieR On-line Data Catalog, 2007.






\end{thebibliography}
\end{document}